\documentclass{iaus}
\usepackage{graphicx}

\begin{document}

%**********************************

\title{Astrometric Solar-System Anomalies}

\author[John D. Anderson \& Michael Martin Nieto]   
{John D. Anderson$^1$ \and Michael Martin Nieto$^2$}

\affiliation{$^1$Jet Propulsion Laboratory (Retired)\\121 S. Wilson Ave., Pasadena, CA 91106-3017 U.S.A. \\ email: {\tt jdandy@earthlink.net}
\\[\affilskip]
$^2$Theoretical Division (MS-B285), Los Alamos National Laboratory\\
Los Alamos, New Mexico 87645 U.S.A. \\ email: {\tt mmn@lanl.gov}
}

\pubyear{2009}
\volume{261}  %% insert here IAU Symposium No.
\pagerange{1--8}
% \date{?? and in revised form ??}
\setcounter{page}{1}
\jname{Relativity in Fundamental Astronomy:\\
Dynamics, Reference Frames, and Data Analysis}
\editors{Sergei Klioner, P. Kenneth Seidelmann \& Michael Soffel}

%***********************************************

\maketitle

\begin{abstract}
There are at least four unexplained anomalies connected with astrometric data. Perhaps the most disturbing is the fact that when a spacecraft on a flyby trajectory approaches the Earth within 2000 km or less, it often experiences a change in total orbital energy per unit mass. Next, a secular change in the astronomical unit AU is definitely a concern. It is reportedly increasing by about 15 cm yr$^{-1}$. The other two anomalies are perhaps less disturbing because of known sources of nongravitational acceleration. The first is an apparent slowing of the two Pioneer spacecraft as they exit the solar system in opposite directions. Some astronomers and physicists, including us, are convinced this effect is of concern, but many others are convinced it is produced by a nearly identical thermal emission from both spacecraft, in a direction away from the Sun, thereby producing acceleration toward the Sun. The fourth anomaly is a measured increase in the eccentricity of the Moon's orbit. Here again, an increase is expected from tidal friction in both the Earth and Moon. However, there is a reported unexplained increase that is significant at the three-sigma level. It is prudent to suspect that all four anomalies have mundane explanations, or that one or more anomalies are a result of systematic error. Yet they might eventually be explained by new physics.  For example, a slightly modified theory of gravitation is not ruled out, perhaps analogous to Einstein's 1916 explanation for the excess precession of Mercury's perihelion.

%Keyword1, keyword2, keyword3, etc.}
\end{abstract}

%****************************************************

\firstsection % if your document starts with a section,
\section{Earth flyby anomaly}

The first of the four anomalies considered here is a change in orbital energy for spacecraft that fly past the Earth on approximately hyperbolic trajectories (\cite[Anderson et al. 2008]{Anderson_etal08}). By means of a close flyby of a planet, it is possible to increase or decrease a spacecraft's heliocentric orbital velocity far beyond the capability of any chemical propulsion system (see for example \cite[Flandro 1966]{Flandro66} and \cite[Wiesel 1989]{Wiesel89}). It has been known for over a century that when a small body encounters a planet in the solar system, the orbital parameters of the small body with respect to the Sun will change. This is related to Tisserand's criterion for the identification of comets (\cite[Danby 1988]{Danby88}).

%The interplanetary gravity-assist technique was observed for the first time 
%on Pioneer 10 (see Sec. \ref{pio}) when it encountered Jupiter on 1973-Dec-04.  In the inner solar system it was used  
%on the NASA Mariner 10 Mission to Mercury, both for a gravity assist by Venus on 1974-Feb-05 for an arrival at Mercury on 1974-Mar-29, and for a gravity %assist during the first Mercury flyby such that two more useful Mercury flybys were achieved (\cite[Dunne \& Burgess 1978]{DunneBurgess78}). 

During a gravity assist, which is now routine for interplanetary missions, the orbital energy with respect to the planet is conserved. Therefore, if there is an observed energy increase or decrease with respect to the planet during the flyby, it is considered anomalous (\cite[Anderson et al. 2007]{Anderson_etal07}).

Unfortunately, it is practically impossible to detect a small energy change with planetary flybys, both because an energy change is difficult to separate from errors in the planet's gravity field and because of the unfavorable Doppler tracking geometry of a distant planet. The more favorable geometry of an Earth flyby is needed. Also the Earth's gravity field is well known from the GRACE mission (\cite[Tapley et al. 2004]{Tapley_etal04}). Earth's gravity is not a significant source of systematic error for the flyby orbit determination (\cite[Anderson et al. 2008]{Anderson_etal08}).

The flyby anomaly was originally detected in radio Doppler data from the first of two Earth flybys by the Galileo spacecraft (for a description of the mission see \cite[Russell 1992]{Russell92}). After launch on 1989-Oct-18, the spacecraft made one flyby of Venus on 1990-Feb-10, and subsequently two flybys of Earth on 1990-Dec-08 and two years later on 1992-Dec-08. The spacecraft arrived at Jupiter on 1995-Dec-07.

Without these planetary gravity assists, a propulsion maneuver of 9 km s$^{-1}$ would have been needed to get from low Earth orbit to Jupiter. With them, the Galileo spacecraft left low Earth orbit with a maneuver of only 4 km s$^{-1}$. The first Earth flyby occurred at an altitude of 960 km. The second, which occurred at an altitude of 303 km, was affected by atmospheric drag, and therefore it was difficult to obtain an unambiguous measurement of an anomalous energy change on the order of a few mm~s$^{-1}$.

\begin{figure}[h]
% \vspace*{-2.0 cm}
\begin{center}
 \includegraphics[width=3.4in]{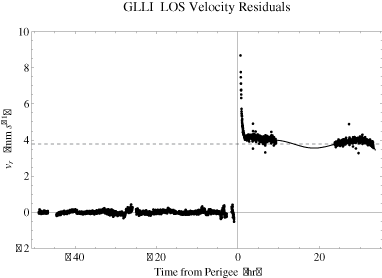} 
% \vspace*{-1.0 cm}
\caption{Doppler residuals (observed minus computed) converted to units of line of sight (LOS) velocity about a fit to the pre-perigee Doppler data, and the failure of this fit to predict the post perigee data. The mean offset in the post-perigee data approaches 3.78 mm s$^{-1}$, as shown by the dashed line. The solid line connecting the post-perigee data represents an eighth degree fitting polynomial to data after perigee plus 2.30 hours. The time of perigee is 1990-Dec-08 20:34:34.40 UTC.}
   \label{fig1}
\end{center}
\end{figure}

The anomalistic nature of the flyby is demonstrated by Fig.\,\ref{fig1}. The pre-perigee fit produces residuals which are distributed about a zero mean with a standard error of 0.087 mm s$^{-1}$. However, when the pre-perigee fit is extrapolated to the post-perigee data, there is a clear asymptotic bias of 3.78 mm s$^{-1}$ in the residuals. Further, the data immediately after perigee indicates that there is perhaps an anomalous acceleration acting on the spacecraft from perigee plus 2253 s, the first data point after perigee, to about 10 hr, the start of the asymptotic bias.  (A discussion of these residuals and how they were obtained can be found in \cite[Antreasian \& Guinn 1998]{AntreasianGuinn98}.)

%*****************************************

%**************************************************

\begin{figure}[h]
% \vspace*{-2.0 cm}
\begin{center}
 \includegraphics[width=3.4in]{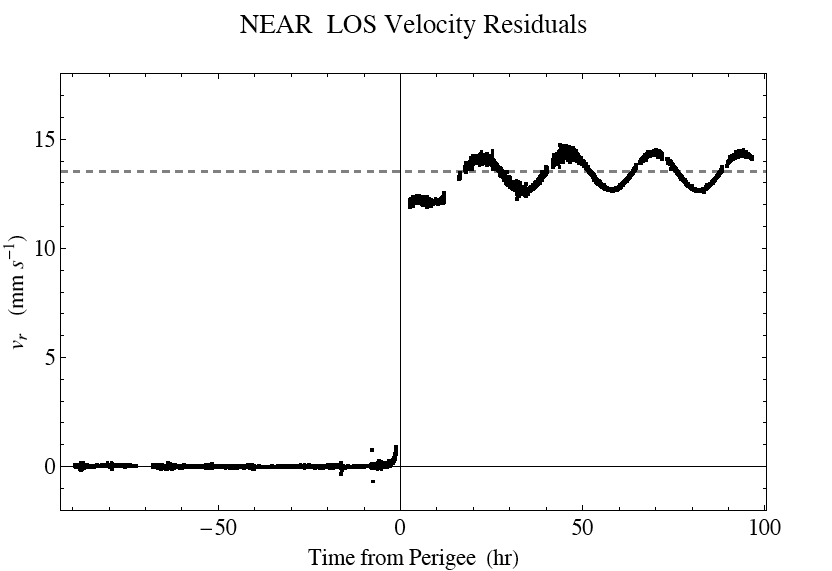}
% \includegraphics[width=3.4in]{NEAR.eps} 
% \vspace*{-1.0 cm}
\caption{Similar to Fig.\,\ref{fig1} but for the NEAR Doppler residuals. The mean offset in the post-perigee data approaches 13.51 mm s$^{-1}$, as shown by the dashed line. The post-perigee data start at perigee plus 2.51 hours. The time of perigee is 1998-Jan-23 07:22:55.60 UTC.}
   \label{fig2}
\end{center}
\end{figure}

A similar but larger effect was observed during an Earth flyby by the Near Earth Asteroid Rendezvous (NEAR) spacecraft. The spacecraft took four years after launch to reach the asteroid (433) Eros in February 2000 (\cite[Dunham et al. 2005]{Dunham_etal05}). For the Earth gravity assist in January 1998, the pre-perigee Doppler data can be fit with a residual standard error of 0.028 mm s$^{-1}$. Note that the residuals are smaller for NEAR with its Doppler tracking in the X-Band at about 8.0 GHz, as opposed to Galileo in the S-Band at about 2.3 GHz. Scattering of the two-way radio signal by free ionospheric electrons is less at the higher frequency, although systematic and random effects from atmospheric refraction limit the X-Band tracking accuracy. Nevertheless, the post-perigee residuals (\cite[Antreasian \& Guinn 1998]{AntreasianGuinn98}) show a clear asymptotic bias of 13.51 mm s$^{-1}$ (see Fig.\,\ref{fig2}). There is also some evidence from Fig.\,\ref{fig2} that an anomalistic acceleration might be acting over perhaps plus and minus 10 hours of perigee.

%******************************************************

%*****************************************************

The anomalistic bias can also be demonstrated for both GLLI and NEAR by fitting the post-perigee data and using that fit to predict the pre-perigee residuals (\cite[Anderson et al. 2008]{Anderson_etal08}). For both spacecraft, the two pre- and post-perigee fits are consistent with the same velocity increases shown in Fig.\,\ref{fig1} and Fig.\,\ref{fig2}.

Earth flybys by the Cassini spacecraft on 1999-Aug-18 and the Stardust spacecraft in January 2001 yielded little or no information on the flyby anomaly. Both spacecraft were affected by thruster firings which masked any anomalous velocity change. However, on 2005-Mar-04 the Rosetta spacecraft swung by Earth on its first flyby and an anomalous energy gain was once again observed. Rosetta is an ESA mission with space navigation by the European Space Operations Center (ESOC). As such it provides an independent analysis at ESOC for both ESA and NASA tracking data for Rosetta (\cite[Morley \& Budnik 2006]{MorleyBudnik06}). The Rosetta anomaly was confirmed independently at JPL with an asymptotic velocity increase of (1.80 $\pm$ 0.03) mm s$^{-1}$ (\cite[Anderson et al. 2008]{anderson_etal08}). Similar data analysis by \cite{Anderson_etal08} yielded slightly different velocity changes than indicated by Fig.\,\ref{fig1} and Fig.\,\ref{fig2} but with error bars. The best estimates are (3.92 $\pm$ 0.03) mm s$^{-1}$ for GLLI and (13.46 $\pm$ 0.01) mm s$^{-1}$ for NEAR.
Rosetta swung by the Earth again on 2007-Nov-13 (RosettaII), but this time no anomaly was reported.

There is most likely a distance dependence to the anomaly. The net velocity increase is 3.9 mm s$^{-1}$ for the Galileo spacecraft at a closest approach of 960 km, 13.5 mm s$^{-1}$ for the NEAR spacecraft at  539 km,  and 1.8 mm s$^{-1}$ for the Rosetta spacecraft at 1956 km. The altitude of RosettaII is 5322 km, perhaps too high for a detection of the anomaly. A third Rosetta Earth swing-by (RosettaIII) is scheduled for 2009-Nov-13 at a more favorable altitude of 2483 km. This third gravity assist, which possibly could reveal the anomaly, will place Rosetta on a trajectory to rendezvous with Comet 67P/Churyumov--Gerasimenko on 2014-May-22 and a lander will be placed on the comet on 2014-Nov-10. The spacecraft bus will orbit the comet and escort it around the Sun until December 2015, when the comet will be at a heliocentric distance of about one AU.

Indeed there is a distance-independent phenomenological formula that models the anomaly quite accurately, at least for flybys at an altitude of 2000 km or less, be that fortuitous or not (\cite[Anderson et al. 2008]{anderson_etal08}). The percentage change in the excess velocity at infinity $v_\infty$ is given by
\begin{eqnarray}
\frac{\Delta v_\infty}{v_\infty} &=& K(\cos \delta_i - \cos \delta_f), 
     \label{formula} \\
K &=& \frac{2\omega_{\oplus} R_{\oplus}}{c} = 3.099 \times 10^{-6},
\end{eqnarray}   
where $\delta_{\{i,f\}}$  are the initial (ingoing) and final (outgoing) 
declination angles given by 
\begin{equation}
\sin \delta_{\{i,f\}} = \sin I \cos \left( \omega \mp \psi  \right).  
\end{equation}
The parameter $\omega_{\oplus}$ is the Earth's angular velocity of rotation, $R_{\oplus}$ is the Earth's mean radius, and $c$ is the velocity of light.  

The angle $\psi$ is one half the total bending angle in the flyby trajectory, $I$ is the 
osculating orbital inclination to the equator of date, and $\omega$ is the 
osculating argument of the perigee measured along the orbit from the equator of 
date. The angle $\psi$ is related to the osculating eccentricity $e$ by
\begin{equation}
\sin \psi =\frac{1}{e}
\end{equation}
Alternatively, the total bending angle $2 \psi$ can be obtained as the angle 
between the asymptotic ingoing and outgoing velocity vectors.

%***********************************************

\section{Increase in the Astronomical Unit}
Radar ranging and spacecraft radio ranging to the inner planets result in a measurement of the AU to an accuracy of 3 m, or a percentage error of $2\times10^{-11}$, making it the most accurately determined constant in all of astronomy (\cite{Pitjeva07}, \cite{PitjevaStandish09}). In SI units the AU can be expressed by the constant $A$, or as the number of meters or seconds in one AU. The two SI units are interchangeable by means of the defining constant $c$, the speed of light in units m s$^{-1}$. In this form, and in combination with the IAU definition of the AU (Resolution No. 10 1976\footnote{http://www.iau.org/static/resolutions/IAU1976\_French.pdf}), there is an equivalence between the AU and the mass of the Sun $M_S$  given by
\begin{equation}
G M_S \equiv k^2 A^3,
\label{GMAU}
\end{equation}
where $G$ is the gravitational constant and $k$ is Gauss' constant. 

According to IAU Resolution No. 10, $k$ is exactly equal to $0.01720209895~AU^{3/2}~d^{-1}$, similar to $c$ exactly equal to $299792458~m~s^{-1}$. The value of the AU is connected to the ranging observations by the time unit used for the time delay of a radar signal or a modulated spacecraft radio carrier wave, ideally the SI second, or equivalently the day $d$ of 86400 s. The extraordinary accuracy in the AU is based on Earth-Mars spacecraft ranging data over an interval from the first Viking Lander on Mars in 1976 and continuing with Viking from 1976 to 1982, Pathfinder P (1997), MGS from 1998 to 2003, and Odyssey from 2002 to 2008 (\cite{Pitjeva09a, Pitjeva09b}). In practice the AU is measured in units of Coordinated Universal Time (UTC), the time scale used by the Deep Space Network (DSN) in their frequency and timing system. Therefore the AU is given in SI seconds as determined by International Atomic Time TAI (\cite{Moyer03}). The fitting models for the JPL ephemeris and for the IAA-RAS ephemeris (\cite{PitjevaStandish09}) are relativistically consistent with ranging measurements in units of SI seconds. It seems that we really do know the AU to ($149597870700~\pm~3$)~m (\cite{PitjevaStandish09}).

For purposes of deciding whether a measurement of a change in the AU is feasible, we simulate Earth-Mars ranging at a 40-day sample interval over a 27-year observing interval starting on 1976-July-01, for a total of 248 simulated normal points. We approximate the tracking geometry by means of a Newtonian integration of a four-body system consisting of the Sun, the Earth-Moon barycenter, the Mars barycenter, and the Jupiter barycenter, all treated as point masses. The initial conditions of the Earth and Mars are adjusted to give a best fit to the distance between the Earth-Moon barycenter and the Mars barycenter, as given by DE405. The rms error in this best fit is $2.6~\times~10^{-5}$ AU, which is unacceptable as a fitting model, but sufficient for a covariance analysis. In the real analysis (\cite{Pitjeva09a, Pitjeva09b}) the ranging data are represented by hundreds of parameters, only one of which is the AU. 

The parameters for our covariance analysis consist of the 12 state variables for Earth and Mars, expressed as the Cartesian initial conditions at the July 1976 epoch, plus two parameters ($k_1,k_2$) for $G M_S$ as given by $k^2 \left[1 + k_1 + k_2 (t - \bar{t}) \right]$ in units of AU$^{3} d^{-2}$.  This is the most direct way to express a bias in the AU and its secular time variation as a Newtonian perturbation. \footnote{The AU is not determined in ephemeris software by means of this physical approach (see Pitjeva (2007), Pitjeva (2009a) and Pitjeva (2009b) for details).} The masses of the three planetary systems are constant at their DE405 values, and the initial conditions of the Jupiter system are not included in the covariance matrix, which makes it a 14$\times$14 matrix. The rank of this matrix is actually 12. The mean Earth orbit defines the reference plane for the other orbits. Hence there are only four Earth elements that can be inferred from the data. A singular value decomposition (SVD) of the 14$\times$14 matrix can be obtained and its pseudo inverse can be interpreted as the covariance matrix on the 14 parameters (\cite{LawsonHanson74}). Actually all the information on $k_1$ and $k_2$ is obtained by the 8th singular value, so a rank 9 pseudo inverse is more than sufficient for a study of the AU and its time variation. The mean time $\bar{t}$ is introduced into the secular variation in $G M_S$ such that $k_1$ and $k_2$ are uncorrelated. This mean time is 13.5 yr for the simulation, but in the real analysis it should be taken as the mean of all the observation times.

%The tracking geometry for Earth-Mars ranging is shown in Fig.~\ref{MarsOrbit} in an Earth-Moon  barycentric coordinate system rotating at the Earth's %mean orbital period of 365.25636 d. In this system the Sun's orbit is a small ellipse caused by an Earth eccentricity of 0.0167.

%******************************************************

%\begin{figure}[h]
% \vspace*{-2.0 cm}
%\begin{center}
 %\includegraphics[width=3.4in]{MarsOrbit.eps} 
% \vspace*{-1.0 cm}
%\caption{The highly variable Earth-Mars distance over a time interval of 27 yr is accentuated by a relatively large orbital eccentricity for Mars of 0.0934. In a coordinate system rotating with the Earth, the Mars orbit is viewed as clockwise at the Earth-Mars synodic period of 2.135 yr. The beginning of this synodic orbit approaching a near-maximum conjunction at the epoch of 1976-Jul-01 is indicated by a  large filled circle, as is its end approaching a near-minimum opposition 27 years later.}
 %  \label{MarsOrbit}
%\end{center}
%\end{figure}

%*****************************************************
 
Taking account of the factor of three in Eq.~\ref{GMAU}, we normalize the result of the covariance analysis to a standard error in the AU of 3.0 m, represented by $k_1$ in the rank 12 matrix. The corresponding rank 9 standard error, where it is assumed that all the remaining five singular values are perfectly known, is 2.5 m. The corresponding error in the secular variation represented by $k_2$ is 2.9 cm yr$^{-1}$ for full rank 12 and 2.7 cm yr$^{-1}$ for rank 9. 

We conclude that at least the uncertainty part of the reported increase in the AU (\cite{KrasinskyBrumberg04}) of (15 $\pm$ 4) cm yr$^{-1}$ is reasonable. Any future work should be focused on checking the actual mean value of the secular increase and perhaps refining it. It is unlikely that its error bar can be decreased below 3.0 cm yr$^{-1}$ with existing Earth-Mars ranging data. However, if the error in the AU can be reduced to $\pm$ 1.0 m with confidence, the error in its secular variation could perhaps be reduced to $\pm$ 1.0 cm yr$^{-1}$, with Earth-Mars ranging alone. Other than that, the Cassini spacecraft carries an X-Band ranging transponder (\cite[Kliore 2004]{Kliore04}). Range fixes on Saturn presumably can be obtained for each Cassini orbital period of roughly 14.3 days over an observing interval from July 2004 to July 2009, or for as long as the spacecraft is in orbit about Saturn and ranging data are available. These data are not yet publicly available, but when they are released, we can expect a standard error in each ranging normal point of about 5 m. Spacecraft ranging to Mercury during the MESSENGER and BepiColombo missions could also add additional information for the AU and its secular variation. If the AU is really increasing with time, the planetary orbits by definition (Eq.~\ref{GMAU}) are shrinking and their periods are getting shorter, such that their mean orbital longitudes are increasing quadratically with $t$, the major effect that can be measured with Earth-planet ranging data.

However, rather than increasing, the AU should be decreasing, mainly as a result of loss of mass to solar radiation, and to a much lesser extent to the solar wind. The total solar luminosity is 3.845~$\times~10^{26}$~W (\cite{Livingston99}). This luminosity divided by c$^2$ gives an estimated mass loss of 1.350~$\times~10^{17}$ kg yr$^{-1}$. The total mass of the Sun is 1.989~$\times~10^{30}$ kg (\cite{Livingston99}), so the fractional mass loss is 6.79~$\times~10^{-14}$ yr$^{-1}$. Again with the factor of three from Eq.~\ref{GMAU}, the expected fractional decrease in the AU is 2.26~$\times~10^{-14}$ yr$^{-1}$, or a change in the AU of $-~0.338$~cm~yr$^{-1}$. A change this small is not currently detectable, and it introduces an insignificant bias into the reported measurement of an AU increase (\cite{KrasinskyBrumberg04}). If the reported increase is absorbed into a solar mass increase, and not into a changing gravitational constant G, the inferred solar mass increase is  (6.0~$\pm$~1.6)~$\times~10^{18}$ kg~yr$^{-1}$. This is an unacceptable amount of mass accretion by the Sun each year. It amounts to a fair sized planetary satellite of diameter 140 km and with a density of 2000 kg m$^{-3}$, or to about 40,000 comets with a mean radius of 2000 m. If the reported increase holds up under further scrutiny and additional data analysis, it is indeed anomalous. Meanwhile it is prudent to remain skeptical of any real increase. In our opinion the anomalistic increase lies somewhere in the interval zero to 20 cm yr$^{-1}$, with a low probability that the reported increase is a statistical false alarm.

%*************************************

\section{The Pioneer anomaly} 
\label{pio}

The first missions to fly to deep space were the Pioneers.  By using flybys, heliocentric velocities were obtained that were unfeasible at the time by using only chemical fuels.
Pioneer 10 was launched on 1972-Mar-02 local time.  
It was the first craft launched into deep space and was the first to reach an outer giant planet, Jupiter, on 1973-Dec-04.   
With the Jupiter flyby, Pioneer 10 reached escape velocity from the solar system.
Pioneer 10 has an asymptotic escape velocity from the Sun of 11.322 km s$^{-1}$ (2.388 AU yr$^{-1}$).

Pioneer 11 followed soon after Pioneer 10, with a launch on 1973-Apr-06.  It too cruised  to Jupiter on an approximate heliocentric ellipse.  This time a carefully executed flyby of Jupiter put the craft on a trajectory to encounter Saturn in 1979.  So, on 1974-Dec-02, when Pioneer 11 reached Jupiter, it underwent a Jupiter gravity assist that sent it back inside the solar system to catch up with Saturn on the far side.   It was then still on an ellipse, but a more energetic one.  
Pioneer 11 reached Saturn on 1979-Sept-01.   Then Pioneer 11 embarked on an escape hyperbolic trajectory with an asymptotic escape velocity from the Sun of 10.450 km s$^{-1}$ (2.204 AU yr$^{-1}$)

The Pioneer navigation was carried out at the Jet Propulsion Laboratory.  It used 
NASA's DSN to transmit and obtain the raw radiometric data.  An S-band signal ($\sim$2.11 Ghz) was sent up via a DSN antenna located either at Goldstone, California, outside Madrid, Spain, or outside Canberra, Australia.  On reaching the craft the signal was transponded back with a (240/221) frequency ratio ($\sim$2.29 Ghz), 
and received back at the same station (or at another station if, during the radio round trip, the original station had rotated out of view).  There the signal was compared with 240/221 times the recorded transmitted frequency and any Doppler frequency shift was measured directly by cycle count compared to an atomic clock. 
The processing of the raw cycle count produced a data record of Doppler frequency shift as a function of time, and from this a trajectory was calculated.  This procedure was done iteratively for purposes of converging to a best fit by nonlinear weighted least squares (minimization of the chi squared statistic, see \cite{LawsonHanson74}).

However, to obtain the spacecraft velocity as a function of time from this Doppler shift is not easy.  The codes must include all gravitational and time effects of general relativity to order $(v/c)^2$ and some effects to order $(v/c)^4$.  The ephemerides of the Sun, planets and their large moons as well as the lower mass multipole moments are included.  The positions of the receiving stations and the effects of the tides on the exact positions, the ionosphere, troposphere, and the solar plasma are included.

Given the above tools, precise navigation was possible because, due to a serendipitous stroke of luck, the Pioneers were spin-stabilized. With spin-stabilization the craft are rotated at a rate of  $\sim$(4-7) rpm about the principal moment-of-inertia axis.  Thus, the craft is a gyroscope and attitude maneuvers are needed only when the motions of the Earth and the craft move the Earth from the antenna's line-of-sight.

The Pioneers were chosen to be spin-stabilized because of other engineering decisions.  As the craft would be so distant from the Sun solar power panels would not work.  Therefore these were the first deep spacecraft to use nuclear heat from $^{238}$Pu as a power source in Radioisotope Thermoelectric Generators  (RTGs).  Because of the then unknown effects of 
long-term radiation damage on spacecraft hardware, a choice was made to place the RTGs at the end of long booms.  This placed them away from the craft and thereby avoided most of the radiation that might be transferred to the spacecraft.

Even so, there remained one relatively large effect on this scale that had to be modeled: the solar radiation pressure of the Sun.  This effect is approximately 1/30,000 that of the Sun's gravity on the Pioneers.  It produced an acceleration of  $\sim 20 \times 10^{-8}$ cm s$^{-2}$ on the Pioneer craft at the distance of Saturn.

\begin{figure}[h!] 
    \noindent
    \begin{center}  
 \includegraphics[width=3.4in]{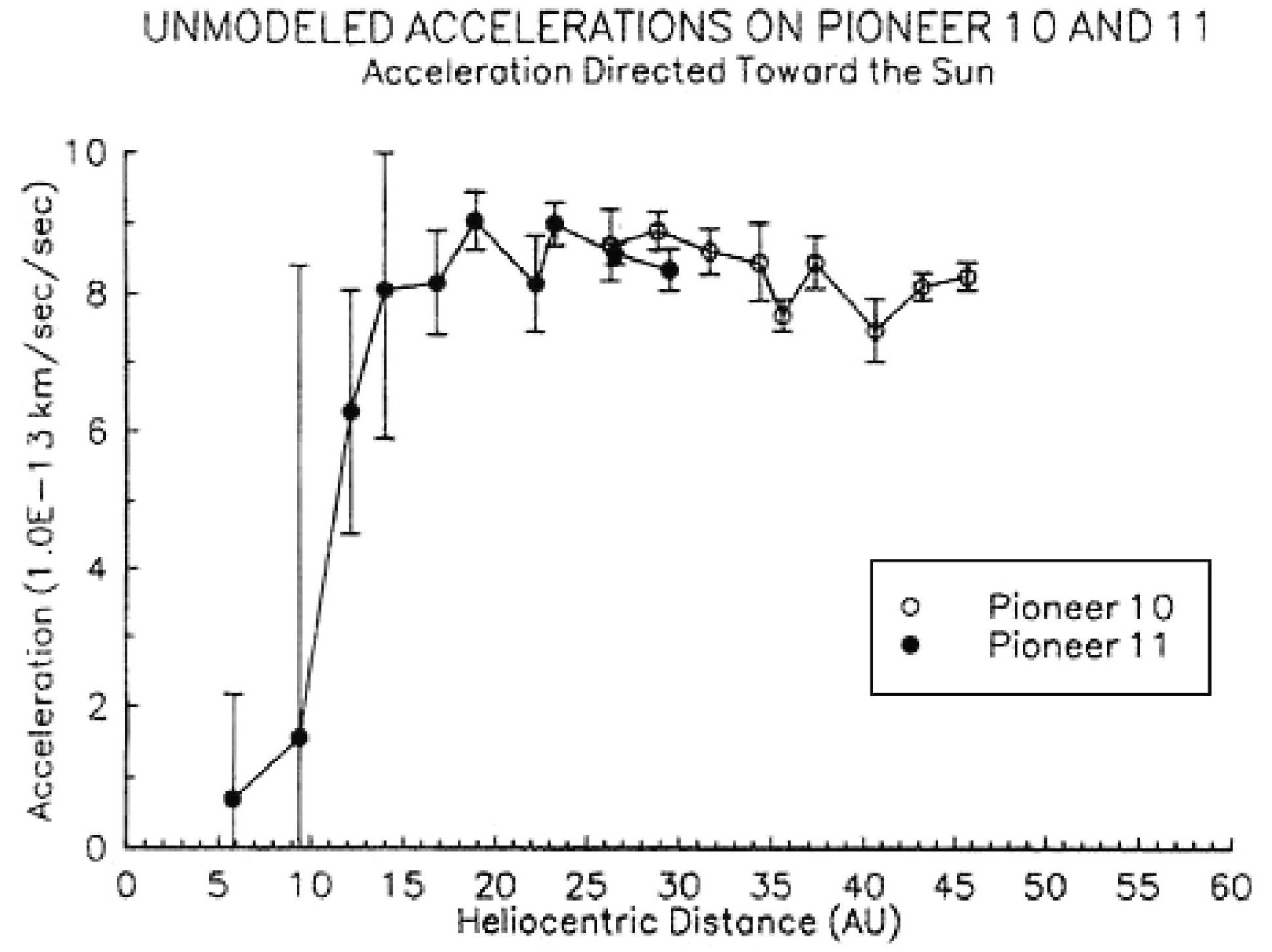}
\caption{A JPL Orbit Determination Program (ODP) plot of the early unmodeled accelerations of Pioneer 10 and Pioneer 11, from about 1981 to 1989 and 1977 to 1989, respectively.   
\label{correlate}}
\end{center}
\end{figure}

After 1976 small time-samples (approximately 6-month to 1-year averages) of the data were periodically analyzed.   At first nothing significant was found,    
But when a similar analysis was done around Pioneer 11 's Saturn flyby, things dramatically changed.  (See the first three data points in Fig. 
\ref{correlate}.)  So people kept following Pioneer 11.  They also started looking more closely at the incoming Pioneer 10 data.

%**************************** 6

%****************************

By 1987 it was clear that an anomalous acceleration appeared to be acting on the craft with a magnitude $\sim 8 \times 10^{-8}$ cm s$^{-2}$, directed approximately towards the Sun.    The effect was a concern, but the effect was small in the scheme of things and did not affect the necessary precision of the navigation.  However, by 1992 it was clear that a more detailed look would be useful.

An announcement was made at a 1994 conference proceedings. The strongest immediate reaction was that the anomaly could well be an artifact of JPL's Orbit Determination Program (ODP), and could not be taken seriously until an independent code had tested it.  So, a team was gathered that included colleagues from The Aerospace Corporation and their independent CHASMP navigation code.  Their result was the same as that obtained by JPL's ODP.

The Pioneer anomaly collaboration's discovery paper appeared in 1998 
(\cite[Anderson et al. 1998]{Anderson_etal98}).
and a detailed analysis appeared in 2002 
(\cite[Anderson et al. 2002]{Anderson_etal02}).
The latter used Pioneer 10 data spanning 1987-Jan-03 to 1998-Jul-22 (when the craft was 40 AU to 70.5 AU from the Sun) and Pioneer  11 data spanning 1987-Jan-05 to 1990-Oct-01 (when Pioneer 11 was 22.4 to 31.7 AU from the Sun). The largest systematics were, indeed, from heat but the final result for the anomaly, is that there is an unmodeled acceleration, directed approximately towards the Sun, of 
\begin{equation}
a_P = (8.74 \pm 1.33) \times 10^{-8}~\mathrm{cm~s^{-2}}.
\end{equation}

Two later and independent analyses of this data obtained similar results.  The conclusion, then, is that this ``Pioneer anomaly" is in the data.  
The question is
(\cite{NietoAnderson07}), ``What is its origin?''

It is tempting to assume that radiant heat must be the cause of the acceleration, since only 63 W of directed power could cause the effect (and much more heat than that is available).  
The heat on the craft ultimately comes from the  Radioisotope Thermoelectric Generators (RTGs), which yield heat from the radioactive decay of $^{238}$Pu.  Before launch,
the four RTGs had a total thermal fuel inventory of 2580 W ($\approx
2070$ W in 2002). 
Of this heat 165 W was converted at launch into electrical power, which decreased down to $\sim70$ W.
So, heat as a mechanism yielding an approximately constant effect remains to be clearly resolved, but detailed studies are underway at JPL.

Indeed, from the beginning we observed that a most likely origin is directed heat radiation 
(\cite[Anderson et al. 1998]{Anderson_etal98}, \cite[Anderson et al. 2002]{Anderson_etal02}).  However, suspecting this likelihood is different from proving it.  Even so, investigation may well ultimately show that heat was a larger effect than originally demonstrated by \cite[Anderson et al. (2002)]{Anderson_etal02}. Their original estimate of the bias from reflected heat amounts to only 6.3\% of the total anomaly. Nevertheless, a three-sigma error in the original estimate could amount to a 25\% thermal effect. We would have difficulty accepting anything larger than this three-sigma limit.

On the other hand, if this is a modification of gravity, it is not universal; i.e., it is not a scale independent force that affects planetary bodies in bound orbits.  
The anomaly could, in principle be i) some modification of gravity, ii) drag from dark matter, or a modification of inertia, or iii) a light acceleration  
(\cite{NietoAnderson07});

Future study of the anomaly may determine which, if any, of these proposals are viable.

%*************************************************

\section{Increase in the eccentricity of the Moon's orbit}
A detailed orbital analysis of Lunar Laser Ranging (LLR) data can be found in \cite[Williams \& Boggs (2009)]{WilliamsBoggs09}. A total of 16,941 ranges were analyzed extending from 1970-Mar-16 to 2008-Nov-22. LLR can measure evolutionary changes in the geocentric lunar orbit over this interval of 38.7 years. Changes in both the  mean orbital motion and eccentricity are
observed. While the mean motion and semi-major axis rates of the lunar orbit are consistent with physical models for dissipation in Earth and Moon, LLR orbital solutions consistently reveal an
anomalous secular eccentricity variation. After accounting for tides on the Earth that produce an eccentricity change of 1.3~$\times$~10$^{-11}$~yr$^{-1}$ and tides on the Moon that produce a change of -0.6~$\times$~10$^{-11}$~yr$^{-1}$, there is an anomalous rate of (0.9 $\pm$ 0.3)~$\times$~10$^{-11}$~yr$^{-1}$, equivalent to an extra 3.5~mm~yr$^{-1}$ in perigee and apogee distance (\cite{WilliamsBoggs09}). This anomalous eccentricity rate is not understood and it presents a problem, both for a physical understanding of dissipative processes in the interiors of Earth and Moon, and for the modeling of dynamical evolution at the 10$^{-11}$~yr$^{-1}$ level.

%*****************************************************************

%***************************************************

\end{document}